\RequirePackage{fix-cm}
\documentclass{article}                     

\usepackage{graphicx}
\usepackage[title]{appendix}
\usepackage{amsmath}
\usepackage{lineno}
\usepackage{units}
\usepackage[colorlinks=true]{hyperref}
\usepackage[nomarginpar]{geometry}
\usepackage{authblk}
\begin{document}
\title{GeoChemFoam: Direct modelling of multiphase reactive transport in real pore geometries with equilibrium reactions}
%



\author{Julien Maes, Hannah P. Menke}


\affil{Institute of GeoEnergy Engineering, Heriot-Watt University, U.K}         
\maketitle

\begin{abstract}
We present the novel numerical model GeoChemFoam, a multiphase reactive transport solver for simulations on complex pore geometries, including microfluidic devices and micro-CT images. The geochemical model includes bulk and surface equilibrium reactions. Multiphase flow is solved using the Volume-Of-Fluid method and the transport of species is solved using the Continuous Species Transfer method. The reactive transport equations are solved using a sequential Operator Splitting method, with the transport step solved using our OpenFOAM\textsuperscript{\textregistered}-based Computational Fluid Dynamics toolbox, and the reaction step solved using Phreeqc, the US geological survey's geochemical solver. The model is validated by comparison with analytical solutions in 1D and 2D geometries. We then applied the model to simulate multiphase reactive transport in two test pore geometries:  a 3D pore cavity and a 3D micro-CT image of Bentheimer sandstone. In each case, we show the pore-scale simulation results can be used to develop upscaled models that are significantly more accurate than standard macro-scale equilibrium models.
\end{abstract}

\section{Introduction}
\label{intro}
Reactive transport in porous media is an essential field of study with broad ranging applications in a range of industries including oil and gas production, carbon dioxide (CO$_2$) and hydrogen (H$_2$) storage, geothermal energy production, nuclear waste disposal and subsurface contaminant transport \cite{2005-Steefel}.
These processes include fluid flow with inertia and viscous effects, advective species transport, molecular diffusion, and chemical reactions. In addition, multiple fluid phases are often present, resulting in capillary effects and interface transfer. For such complex systems, analytical solutions are restricted to very simple geometries and flow conditions \cite{1911-Hadamard,2006-Coutelieris}.
These limitations in model complexity result in the use of experiments to investigate more complex physics with small analogue systems such as core samples \cite{2014-Menke,2017-Menke} or reactive micromodels \cite{2017-Soulaine,2020-Poonoosamy}. However, experimental studies are often time-consuming, limited in size, and hard to control. In addition, reactive transport experiments on core samples are always destructive, and since no two cores are the same, they cannot be repeated on identical natural pore structures. These studies are thus often coupled with numerical simulations, a powerful tool that can be used both during the design of the experiment to choose optimal conditions, or to augment the experimental data afterwards by providing quantities of interest that may be difficult to measure (e.g. pH) or to explore additional ranges of physical conditions (e.g. pressure, temperature) \cite{2021-Soulaine}.

While numerical modelling of multiphase flow \cite{2019b-Pavuluri,2015-Ferrari,2019-Zhao} and single-phase reactive transport \cite{2009-Szymczak,2017-Soulaine,2020-Oliveira} in pore-scale geometries have been extensively investigated independently, few studies have attempted to model the coupling between the two. 
Raoof et al. \cite{2013-Raoof} used a pore network model to simulate reactive transport in variably saturated porous media. However, the pore network approach introduces restrictions on the transport regime and reactive surface area calculations. Chen et al. \cite{2018-Chen} employed the Lattice Boltzmann Method (LBM) to model multiphase reactive transport, with an interfacial reaction treatment rather than a direct modelling of interfacial conditions. Although this method has been used successfully in several studies \cite{2017-Chen,2019-CHen}, the LBM method has difficulty modelling the full range of regimes that occur during multiphase flow \cite{2019-Zhao} and reactive transport \cite{2020-Molins}. Aziz et al. \cite{2019-Aziz} investigated wettability alteration during low-salinity flooding using a non-reactive multiphase transport solver based on Direct Numerical Simulation (DNS). However, the model was restricted to transport in the aqueous phase with an immobile non-aqueous phase and no interfacial conditions. None of these studies include accurate modelling of interfacial conditions with phase transfer.

Recent advances in the development of DNS of multiphase transport have enabled accurate modelling of interfacial transfer. Haroun et al. \cite{2010-Haroun} introduced the single-field approach to model species transport in multiphase systems with interfacial conditions. Their method is based on the Volume-Of-Fluid (VOF) method \cite{1981-Hirt}, where the interface between the two fluids is captured using an indicator function, which is a phase volume fraction. Although other
methods such as level-set \cite{2018-Gibou,2019-Luo} can provide a more accurate description of the sharp interface, the VOF method is attractive due to its accuracy of mass conservation and adaptability to more complex physics.  
Marschall et al. \cite{2012-Marschall} developed Haroun's single-field approach into a versatile and precise method for multiphase transport during bubbly flow labelled Continuous Species Transport (CST). This method was extended to problems with moving contact lines by Graveleau et al. \cite{2017-Graveleau} and later improved by Maes and Soulaine \cite{2018-Maes} with the introduction of interface compression. The CST method was then used to model multiphase reactive transport during low-salinity flooding \cite{2017-Maes} and mineral dissolution with CO$_2$ production in shale formations \cite{2018-Soulaine,2019-Soulaine}. Finally, the model was extended to include local volume changes induced by interface transfer for simulating dissolution of CO$_2$ bubbles in liquid \cite{2018b-Maes,2020-Maes,2020-Patsoukis}.

The objective of this paper is to present our extended model that includes multiphase reactive transport with equilibrium reactions both in the water phase and at the surface of the solid. The fully-coupled multiphase reactive transport model is presented in section \ref{sec:1} and validated in section \ref{sec:2}. In particular, we show that precise representation of interfacial conditions is essential for accurate and robust modelling of reactive transport, even when the species only exist in one phase, demonstrating that the CST method can be used for reactive transport,  unlike the model presented in \cite{2019-Aziz}. Finally, we present the simulation and upscaling of  reactive transport with two model test cases:  (1) First, we simulate carbonic acid formation during dissolution of a CO$_2$ gas bubble in a 3D pore cavity and then (2)  we introduce the first results of a multiphase reactive transport simulation on a real 3D pore space with injection of a CaCl solution into a micro-CT image of Bentheimer sandstone.

\section{Model description}
\label{sec:1}

\subsection{Geochemical model}

We consider a multiphase system with a reactive phase $p$ in a chemical model that includes $N_c$ and $N_s$ bulk and surface components, with $N_x$ and $N_y$ bulk and surface equilibrium reactions.  Since the species are at chemical equilibrium, it is possible to partition the system into $\overline{N}_c=N_c-N_x$ and $\overline{N}_s=N_s-N_y$, the primary bulk and surface species (i.e. species with independent concentrations), and $N_x$ and $N_y$, the secondary bulk and surface species \cite{2015-Steefel}. The equilibrium chemical reactions between the primary and secondary species can be written as
\begin{linenomath}
\begin{align}\label{reac}
        A_i \rightleftharpoons \sum_{j=1}^{\overline{N}_c} \nu_{ij}A_j, && S_n \rightleftharpoons S_m +\sum_{j=1}^{\overline{N}_c}\nu_{nj}A_j,
\end{align}
\end{linenomath}
where $A_j$ and $A_i$ are the chemical formulas of the primary
and secondary species in the bulk phase, $S_m$ and $S_n$ are the chemical formulas of the primary and secondary species on the solid surface, and $\nu_{ij}$ and $\nu_{nj}$ are the stoichiometric coefficients. Note that on the solid surface, one secondary species is associated to one primary species only. Each equilibrium reaction provides an algebraic
link between the primary and secondary species via the law
of mass actions
\begin{linenomath}
\begin{align}\label{Eq:massaction}
        a_{i,p} =K_i^{-1} \prod_{j=1}^{\overline{N}_c} a_{j,p}^{\nu_{ij}}, && \omega_n = \omega_m K_n^{-1} \prod_{j=1}^{\overline{N}_c}a_{j,p}^{\nu_{nj}},
\end{align}
\end{linenomath}
where $a_{j,p}$ and $a_{i,p}$ are the activities of primary species $j$ and secondary species $j$, $\omega_m$ and $\omega_n$ are the activity of the primary surface species $m$ and secondary surface species $n$, and $K_i$ and $K_n$ the chemical equilibrium constants. We assume that the activity of a species $k$ in phase $p$ is equal to
\begin{linenomath}
\begin{align}
    a_{k,p}=\gamma_{k,p}\frac{c_{k,p}}{c_0}
\end{align}
\end{linenomath}
where $\gamma_{k,p}$ is the activity coefficient of species $k$ (primary or secondary) , $c_{k,p}$ is its concentration (kmol/m$^3$) in phase $p$ and $C_0=1$ kmol/m$^3$ is the standard activity. The activity $\omega_l$ of a surface species $S_l$ (primary or secondary) is equal to its mole fraction on the corresponding surface, i.e. over all surface components which share the same primary species $S_m$. 
For each primary bulk species $j$, we also define the total concentration $\psi_{j,p}$ in phase $p$, which is the quantity conserved during chemical reactions, and can be written as
\begin{equation}\label{Eq::Psi}
    \psi_{j,p} = c_{j,p} + \sum_{i=1}^{N_x}\nu_{ij}c_{i,p} + \sum_{n=1}^{N_y}\nu_{nj}\omega_n \Gamma A_s,
\end{equation}
where $\Gamma$ is the site density (kmol/m$^2$) and $A_s$ is the specific surface area (m$^2$/m$^3$) of the solid which, at the pore-scale, is calculated from the mesh.  

For surface reactions, the apparent stability constant $K_n$ is different from the intrinsic constant $K^{i}_n$ due to the surface charge $q$
\begin{equation}\label{sigma}
q=F\sum_{n=1}^{N_s} v_{n}\omega_n\Gamma,
\end{equation}
where $v_{n}$ is the charge of the surface species $n$ and $F$ ($=9.649\times10^{7}$ C/kmol) is the Faraday constant. The double-layer surface potential $\varphi$ is related to the surface charge by the Grahame equation \cite{1985-Israelachivili}
\begin{equation}\label{Grahame}
q^2=8000\epsilon\epsilon_0RTI\left[\sinh\left(\frac{F\varphi}{2RT}\right)\right]^2,
\end{equation}
where $\epsilon$ ($=78.41$ at 25$^o$ C) is the dielectric constant of pure water, $\epsilon_0$ ($=8.854\times10^{-12}$ C/V/m) is the vacuum permittivity, $I$ (kmol/m$^3$) is the ionic strength of the electrolyte solution, $R$ ($=8.314$ kJ/kmol/K) is the ideal gas constant and $T$ is the temperature.
The relationship between $K_n$ and $K^i_n$ is given by \cite{1985-Israelachivili}
\begin{equation}\label{Eq:Ki}
K_n = K^i_{n}\exp\left(-\frac{Z_nF\varphi}{RT}\right),
\end{equation}
where $Z_n$ is the net change of surface charge induced by the reaction. In this work, activity coefficients, ionic strength, surface charge, surface potential, and chemical equilibrium constants are calculated within Phreeqc \cite{2013-Pakhurst}.

\subsection{Multiphase flow model:VOF}
In this study, the system includes two phases: the aqueous phase (phase 1) and a non-aqueous phase (phase 2), that can be either a gas or a liquid phase. In the VOF method, the interface between the two fluids is tracked using indicator functions $\alpha_1$ and $\alpha_2$, where $\alpha_2=1-\alpha_1$, which are equal to the volume fractions of each phase in each grid cell. The density and viscosity of the fluid in each cell are expressed using their single-field values
\begin{linenomath}
\begin{align}
    \rho = \alpha_1\rho_1 + \alpha_2\rho_2, \\
    \mu = \alpha_1\rho_1 + \alpha_2\rho_2,
\end{align}
\end{linenomath}
where $\rho_p$ (kg/m$^3$) and $\mu_p$ (Pa.s) are the density and viscosity of phase $p$. Similarly, the velocity and pressure in the domain are expressed in term of the single-field variables
\begin{linenomath}
\begin{align}
    \mathbf{u} = \alpha_1\mathbf{u}_1 + \alpha_2\mathbf{u}_2, \\
    p = \alpha_1p_1 + \alpha_2p_2,
\end{align}
\end{linenomath}
where $\mathbf{u}_p$ (m/s) and $p_p$ (Pa) are the velocity and pressure in phase $p$.  Each phase is assumed to be Newtonian and incompressible, and fluid properties are assumed to be constant in each phase (and in particular independent of the phase composition). In this case, the single-field momentum equation \cite{1981-Hirt} can be written as
\begin{equation}\label{Eq:mom}
\frac{\partial \rho \mathbf{u}}{\partial t}+ \nabla\cdot\left(\rho\mathbf{u}\mathbf{u}\right)=-\nabla p
+ \nabla\cdot\left(\mu\left(\nabla \mathbf{u} + \nabla \mathbf{u}^T\right)\right) + \rho \mathbf{g} + \mathbf{f}_{\sigma},
 \end{equation}
where $\mathbf{g}$ (=9.81 m/s$^2$) is the gravity acceleration and $\mathbf{f}_{\sigma}$ is the surface tension force
\begin{equation}
\mathbf{f}_{st} = \sigma \kappa \mathbf{n}_{12}\delta_{12}.
\end{equation}
where $\sigma$ (N/m) is the interfacial tension, $\mathbf{n}_{12}$ is the normal vector to the fluid/fluid interface, going from phase 1 to phase 2, $\kappa=\nabla\cdot\mathbf{n}_{12}$ is the interface curvature and $\delta_{12}$ is a Dirac function located at the interface.
At the triple point fluid/fluid/solid, the interface forms with the normal to the solid surface a contact angle $\theta$ so that
\begin{equation}
 \mathbf{n}_{12} = \cos \theta \mathbf{n}_s + \sin \theta \mathbf{t}_s,
\end{equation}
where $\mathbf{n}_s$ and $\mathbf{t}_s$ are the normal and tangent vectors to the solid surface, respectively \cite{1992-Brackbill}. 
In addition, the single-field continuity equation writes
\begin{equation}\label{Eq:cont}
    \nabla\cdot\mathbf{u}=\dot{m}_{12}\left(\frac{1}{\rho_2}-\frac{1}{\rho_1}\right).
\end{equation}
where $\dot{m}_{12}$ (kg/m$^3$/s) is the rate of mass transfer from phase 1 to phase 2 by solubility, and is calculated after solving the transport equations.
To advect the indicator functions, algebraic VOF methods solve the phase transport equation
\begin{equation}\label{Eq:phaseEq}
\frac{\partial \alpha_1}{\partial t} + \nabla\cdot(\alpha_1 \mathbf{u}) + \nabla\cdot\left(\alpha_1\alpha_2\mathbf{u}_r\right)=-\frac{\dot{m}_{12}}{\rho_1},
\end{equation}
where $\mathbf{u}_r=\mathbf{u}_1-\mathbf{u}_2$ is the relative velocity, which is a consequence of mass and momentum transfer between the phases. Fleckenstein and Bothe \cite{2015-Fleckenstein} showed that $\mathbf{u}_r$ may be neglected even
in the case of very good solubility (e.g. CO$_2$ in water) in order to simplify Eq. (\ref{Eq:phaseEq}). However, to reduce the smearing of the interface induced by numerical diffusion, an artificial compression term can be introduced by replacing $\mathbf{u}_r$ in Eq. (\ref{Eq:phaseEq})
by a compressive velocity $\mathbf{u}_{comp}$ normal to the interface and with an amplitude based on the maximum of the single-field velocity \cite{2002-Rusche}
\begin{equation}\label{compress}
\mathbf{u}_r\equiv \mathbf{u}_{comp} = \mathbf{n}_{12}\left[\min\left(c_{\alpha}\frac{|\Phi_f|}{A_f},\max_f\left(\frac{|\Phi_f|}{A_f}\right)\right)\right],
\end{equation}
where  $c_{\alpha}$ is the compression constant (generally between 0 and 4), $\Phi_f$ is the volumetric flux across a grid cell face $f$, and $A_f$ is the face area. In all our simulations, we choose $c_{\alpha}=1.0$.

In addition to $\dot{m}_{12}$, which will be calculated in the next section, the system requires models for the normal vector to the fluid/fluid interface and the surface tension force for closure. Brackbill \cite{1992-Brackbill} developed an approximation referred to as the Continuous Surface Force (CSF) where $\mathbf{n}_{12}$ is calculated from $\alpha_1$ and $\mathbf{n}_{12}\delta_{12}$ is approximated by
$\nabla \alpha_1$, so that
\begin{linenomath}
\begin{align}
&\mathbf{n}_{12} = \frac{\nabla \alpha_1}{\| \nabla \alpha_1 \|},
&\mathbf{f}^{CSF}_{st} = \sigma \nabla . \left( \frac{\nabla \alpha_1}{\| \nabla \alpha_1 \|} \right) \nabla \alpha_1.
\end{align}
\end{linenomath}

The VOF-CSF method is attractive because of its simplicity. However, many studies \cite{1999-Scardovelli,2015-Abadie} have reported the presence of spurious currents in the
capillary dominated regime that originate from errors in calculating the normal vector and the curvature of the interface. Spurious currents may be mitigated by a combination of smoothing and sharpening of the indicator functions  \cite{2018-Pavuluri}. Although these modifications of the CSF may reduce the magnitude of spurious currents, they do not fully eliminate them. In addition, they can potentially deteriorate contact line dynamics \cite{2019-Pavuluri}. For these reasons, we do not apply any modifications of the CSF method in this work. Spurious currents exist in our simulations, but their impact has been shown in our previous work \cite{2018-Maes,2020-Maes} to be negligible when compared to analytical solutions. Their impact in more complex geometries has yet to be understood and is a target of future research.  However, in the absence of benchmark experimental data it is impossible to quantify their impact and thus for the purposes of this work we assume them to be negligible.

Multiphase flow in pore structures is generally characterised by two dimensionless numbers, the Reynolds number $Re=\rho_1$UL/$\mu_1$ and the capillary number $Ca=\mu_1$U/$\sigma$, where $U$ and $L$ are the reference velocity and length in the domain, respectively. $Re$ describes the ratio of inertial to viscous forces and $Ca$ the ratio of viscous to capillary forces. In this work, we concentrate our investigation to low flow rates, i.e in the creeping flow and capillary dominated regime with $Re<1$ and $Ca<10^{-4}$.

\subsection{Reactive transport model}
In a multiphase system, the chemical species can be present in both fluid phases. The conservation equation is satisfied by the total concentration $\psi_{j,p}$ of a primary species $j$ (Eq. \ref{Eq::Psi}) in phase $p$ with
\begin{equation}\label{Eq:transport}
\frac{\partial \psi_{j,p}}{\partial t}+ \nabla \cdot \left( \psi_{j,i}\mathbf{u}_i \right) = -\nabla\cdot\mathbf{J}_{j,p},
\end{equation}
where $\mathbf{J}_{j,p}$ is the total diffusive flux of primary species $j$ in phase $p$. We assume that the diffusive flux can be modelled using Fick's law
\begin{linenomath}
\begin{align}
    \mathbf{J}_{j,p} = -D_{j,p}\nabla c_{j_p} - \sum_{i=1}^{N_x}\nu_{ij}D_{i,p}\nabla c_{i,p},
\end{align}
\end{linenomath}
where $D_{j,p}$ and $D_{i,p}$ are the molecular diffusion coefficients of the primary and secondary species in phase $p$. This is true for dilute species in a solvent, such as water, and for species in a pure or binary mixture.
Chemical equilibrium in phase $p$ is insured by the law of mass actions (Eq. \ref{Eq:massaction}). At the fluid/fluid interface, the jump conditions are given by the continuity of fluxes and
chemical potentials, the latter described here by Henry's law \cite{1803-Henry},
\begin{linenomath}
\begin{align}\label{interfaceBC1}
&\left(\psi_{j,1}\left(\mathbf{u}_1-\mathbf{w}\right)+\mathbf{J}_{j,1}\right)\cdot \mathbf{n}_{12} = \left(\psi_{j,2}\left(\mathbf{u}_2-\mathbf{w}\right)+\mathbf{J}_{j,2}\right)\cdot \mathbf{n}_{12},\\
& c_{k,2}=H_kc_{k,1}\label{interfaceBC2}
\end{align}
\end{linenomath}
where $H_k$ is the Henry constant of species $k$ (primary or secondary), while the total mass conservation at the interface is defined as
\begin{linenomath}
\begin{align}\label{interfaceBC3}
    \rho_{1}\left(\mathbf{u}_1-\mathbf{w}\right)\cdot \mathbf{n}_{12} = \rho_{2}\left(\mathbf{u}_2-\mathbf{w}\right)\cdot \mathbf{n}_{12}.
\end{align}
\end{linenomath}
The diffusion coefficients in the aqueous phase and Henry's constants used in this paper are summarized in Table \ref{TableDiffu}. All species exist only in the aqueous phase, except for CO$_2$ that can also exist in the gas phase. In this case, the gas phase will be assumed to be pure. Therefore, the diffusion coefficient of all species in the non-aqueous phase can be assumed to be 0. 
\begin{table}[!ht]
\centering
\begin{tabular}{cccccc}
  Ion & D ($10^{-9}$ m$^2$/s) & H (no unit) & Ion & D ($10^{-9}$ m$^2$/s) & H (no unit)  \\[0.1cm]
\hline
H$^+$ & 9.83 & 0 & OH$^-$ & 5.27 & 0 \\[0.1cm]
CO$_3^{2-}$ & 0.955 & 0 & HCO$_3^{-}$ & 1.18 & 0 \\[0.1cm]
Cl$^{-}$ & 2.03 & 0 & Ca$^{+2}$ & 0.79 & 0 \\[0.1cm]
CO$_2$ & 1.6 & 1.25 & & & \\[0.1cm]
\hline
\end{tabular}
\caption{Diffusion coefficient of ions in water (obtained from \cite{1973-Li}). \label{TableDiffu}}
\end{table}

In order to solve reactive transport within the VOF method, the transport equations (Eq. (\ref{Eq:transport})) are integrated over a control volume using volume averaging \cite{2020-Maes}, and the boundary conditions (Eq. (\ref{interfaceBC1}) and Eq. (\ref{interfaceBC2})) are used to eliminate surface integrals arising from the divergence theorem \cite{1998-Whitaker}. Since the boundary conditions depend on the concentration of primary and secondary species, it is difficult to develop an accurate and stable transport solver for the total concentrations $\left(\psi_j\right)_{1\leq j\leq \overline{N}_c}$. Instead, our model solves directly for the concentration of the primary and secondary species, and is based on a sequential non-iterative operator splitting approach \cite{2004-Carrayrou}. The transport step solves for the single-field concentration of species $k$ (primary or secondary)
\begin{linenomath}
\begin{align}
    c_{k}=c_{k.1}\alpha_1+c_{k,2}\alpha_2,
\end{align}
\end{linenomath}
using the CST method \cite{2020-Maes}. The transport step solves the single-field transport equation
\begin{equation}\label{Eq:CST}
\frac{\partial c_{k}}{\partial t}+ \nabla\cdot\left(c_k\mathbf{u}\right) + \nabla\cdot\left(\alpha_1\alpha_2\left(c_{k,1}-c_{k,2}\right)\mathbf{u}_r\right) - \nabla.\left(D_k\nabla c_k-\mathbf{\Phi}_k\right)=0,
\end{equation}
where
\begin{equation}\label{Eq:PhiCSTb}
\mathbf{\Phi}_k=(1-H_k)D_k\frac{c_k}{\alpha_1+H_k\alpha_2}\nabla\alpha_1,
\end{equation}
is the CST flux of species $k$ and
\begin{align}
    D_k= \frac{\alpha_1D_{k,1}+H_k\alpha_2D_{k,2}}{\alpha_1+H_k\alpha_2},
\end{align}
is the single-field diffusion coefficient of species $k$.
At the surface of the solid, the boundary condition for the single-field concentration of species $k$ is defined by \cite{2017-Graveleau}
\begin{equation}
    D_k\nabla c_k-\Phi_k = 0.
\end{equation}
At the end of the transport step, the rate of mass transfer is calculated by \cite{2020-Maes}
\begin{equation}\label{Eq:mx}
    \dot{m}=-\frac{\sum_{1\leq k<N_c}\left(D_k\nabla c_k -\mathbf{\Phi}_k\right)}{1-\alpha_1} \cdot \nabla\alpha_1. 
\end{equation}
After the transport step is completed, the reaction step is calculated using the phase concentrations $\left(c_{k,p}\right)_{1\leq k\leq N_{c}}$, using the law of mass action (Eq. (\ref{Eq:massaction})) and the mass conservation of the primary species defined as 
\begin{equation}
    \frac{\partial\psi_{j,p}}{\partial t}=0.
\end{equation}

In addition to the Reynolds and capillary numbers, multicomponent multiphase transport in pore structures is generally characterised using the species P\'eclet numbers $Pe_j=UL/D_j$. The transport of a species is advection dominated if $Pe_j>1$, and diffusion dominated if $Pe_j<$1.

\subsection{Interface boundary conditions and artificial mass transfer}

One of the objectives of this paper is to demonstrate that an accurate modelling of interface boundary conditions, such as carried out in the CST method, is necessary for robust modelling of multiphase reactive transport because without such modelling artificial mass transfer may arise that can critically damage the chemical equilibrium. This is true even when no interface transfer exists and the species remain in the water phase.

For this we will compare the transport model presented in this paper with the simplified transport model described in Aziz et al. \cite{2019-Aziz} which only solves for the concentration of species in water (Eq. \ref{Eq:transport}). This is achieved by setting the diffusion coefficient in the non-aqueous phase to 0. The single-field equation is defined as 
\begin{linenomath}
\begin{align}\label{Eq:simplified}
    \text{(Simplified model)} && \frac{\partial c_k}{\partial t}+\nabla\cdot\left(c_k\mathbf{u}-D_{k,1}\alpha_1\nabla c_k\right)=0
\end{align}
\end{linenomath}
It is then generally assumed that a sharp interface between $c_k$ and $\alpha_2$ will be obtained due to the absence of diffusion at the fluid/fluid interface. However, there are two sources of interface transfer that are not accounted for in Eq. (\ref{Eq:simplified}). First, at the interface, $0\leq\alpha_1\leq 1$, so the diffusion coefficient is not 0, even though $D_{k,2}=0$. Second, artificial mass transfer can occur due to the interface compression term in Eq. (\ref{Eq:phaseEq}) if no compression is present in Eq. (\ref{Eq:simplified}) \cite{2020-Maes}. We thus demonstrate in section \ref{sec:3-2} how these unaccounted-for sources of artificial mass transfer may damage the numerical solution.  

\subsection{Upscaling}

Upscaling of multiphase transport in porous media is generally conducted in terms of the Darcy velocity $U_p$, defined using Darcy's law
\begin{equation}
    U_p = \frac{K_ak_{rp}}{\mu_p}\nabla\left(P_p-\rho_p\mathbf{g}\right),
\end{equation}
where $P_p$ is the average pressure in phase $p$, $K_a$ is the absolute permeability of the domain and $k_{rp}$ is the relative permeability of phase $p$. Relative permeabilities are often modelled using the Brooks-Corey model \cite{1964-Brooks} 
\begin{equation}
\begin{aligned}
    k_{r1}= \max\left(0.0,k_{r1,max}\left(\frac{S_1-S_{wc}}{1-S_{wc}-S_{nar}}\right)\right)^{n_1}\\
    k_{r2}= \max\left(0.0,k_{r2,max}\left(\frac{S_{2}-S_{nar}}{1-S_{wc}-S_{nar}}\right)\right)^{n_{2}}
    \end{aligned}
\end{equation}
where $S_p$ is the macro-scale phase saturation,
$S_{wc}$ is the critical water saturation, $S_{nar}$ is the residual non-aqueous saturation, $k_{rp,max}$ is the maximum relative permeability of phase $p$, and $n_p$ is the phase Corey index.
The phase saturation $S_p$ can be calculated from a pore-scale simulation using
\begin{equation}\label{Eq:sat}
    S_p=\frac{1}{V}\int_V\alpha_pdV,
\end{equation}
where the integral is calculated over the whole domain $V$.

The phase velocity $U_p$ is related to the total velocity $U_T=U_1+U_{2}$ by the fractional flow function $f_p$, such as $U_p=f_pU_T$. The fractional flow functions can then be calculated from Darcy's law, and we obtain
\begin{equation}
    f_p=\frac{\frac{k_{rp}}{\mu_p}}{\frac{k_{r1}}{\mu_1}+\frac{k_{r2}}{\mu_{2}}}.
\end{equation}

Multiphase reactive transport in porous media is usually upscaled using an equilibrium model \cite{2016-Chang}, for which the phase saturation $S_p$ (Eq. \ref{Eq:sat}) and the phase average concentrations $C_{j,p}$ are defined as
\begin{equation}
C_{j,p}=\frac{1}{S_pV}\int_V\alpha_pc_{j,p}dV,
\end{equation}
and are computed using an equilibrium phase partitioning. To calculate chemical equilibrium between the species in the aqueous phase, species activities are calculated using the phase average concentrations and then the law of mass actions (Eq. (\ref{Eq:massaction})) is applied. However, due to the slow nature of molecular diffusion in water ($D\sim10^{-9}$ m$^2$/s) and the variation in interfacial area due to pore-size heterogeneity (\cite{2018b-Maes}),
the phase distribution is often more accurately predicted using a linear transfer model \cite{2018-Maes}, for which the transfer $M_k$ (kmol/s) of species $k$ from phase 1 to 2 is calculated as 
\begin{equation}\label{Eq:k}
 M_k=\sum_{1 \leq k\leq N_c}\lambda_k A_{12}\left(H_kC_{k,1}-C_{k,2}\right),
\end{equation}
where $\lambda_k$ (m/s) is the mass exchange coefficient and $A_{12}$ is the interfacial area between phase 1 and phase 2, which can be calculated as
\begin{equation}
    A_{12}=\int_V \|\nabla\alpha_1\|dV.
\end{equation}
In addition, equilibrium models usually overpredict the chemical reaction rates \cite{2015-Alhashmi,2020-Jimenez}. Instead a mixing-induced reaction rate is often introduced as
\begin{equation}
    R_i = k_i \left(1-\Omega_i\right),
\end{equation}
where $k_i$ (kmol/m$^3$/s) is the mixing-induced reaction constant and $\Omega_i$ is the saturation index of reaction $i$. For example, for reaction $i$ in Eq. (\ref{reac}) we define the saturation index as
\begin{equation}
    \Omega_i=\frac{K_i a_{i,p}}{\prod_{j=1}^{\overline{N}_c} a_{j,p}^{\nu_{ij}}}.
\end{equation}
We will show in Section \ref{secCO2} how pore-scale modelling can be applied to calculate mixing reaction rates.

\subsection{Implementation}
\begin{figure}[!t]
\begin{center}
\includegraphics[width=0.55\textwidth]{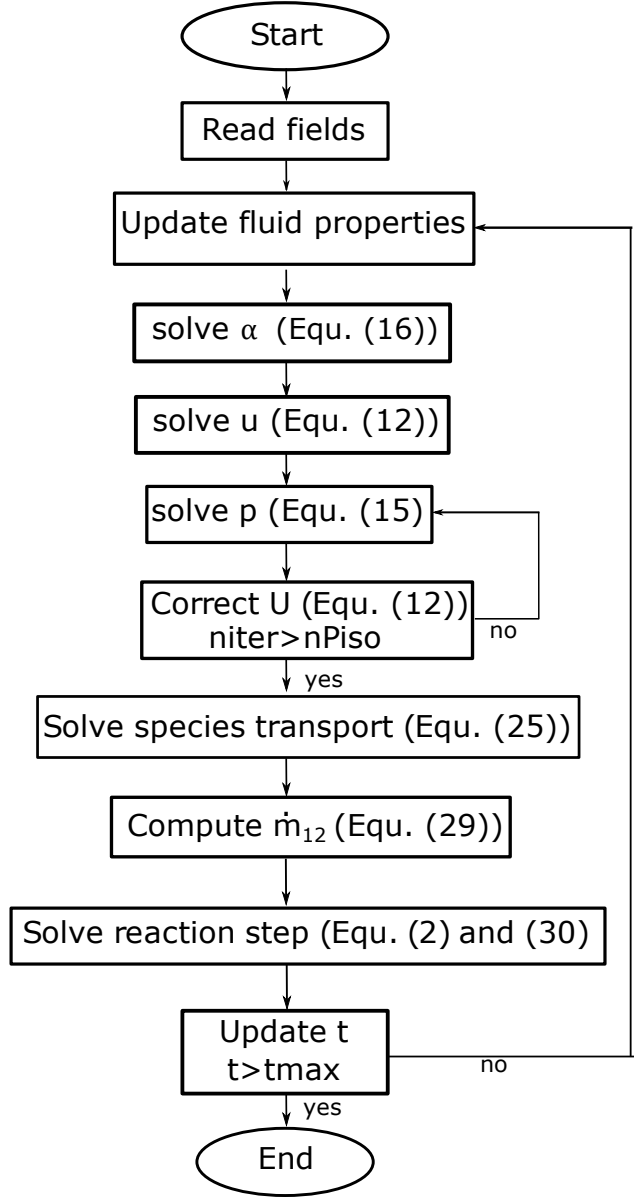}
\caption{Full solution procedure for \textit{interReactiveTransferFoam}\label{fig:solutionProcedure}.}
\end{center}
\end{figure}

The numerical method has been implemented in GeoChemFoam \cite{2020b-Maes}, our reactive transport solver based on OpenFOAM\textsuperscript{\textregistered} \cite{2016-OpenFOAM}. The full code can be downloaded from
\href{www.julienmaes.com}{www.julienmaes.com}. The standard VOF solver of OpenFOAM\textsuperscript{\textregistered}, so-called \textit{interFoam}, has been extended for this purpose into another solver called \textit{interReactiveTransferFoam}.
The full solution procedure is presented in Fig. \ref{fig:solutionProcedure}.

\textit{interFoam} solves the system formed by Eq. (\ref{Eq:cont}), (\ref{Eq:phaseEq}) and (\ref{Eq:mom}) on a collocated Eulerian grid. A pressure equation is obtained by combining the continuity (Eq. (\ref{Eq:cont})) and momentum (Eq. (\ref{Eq:mom})) equations.
These equations are then solved with a predictor-corrector strategy based on the Pressure Implicit Splitting Operator (PISO) algorithm \cite{1985-Issa}. Three iterations of the PISO loop are used to stabilise the system. 
An explicit formulation is used to treat the coupling between the phase distribution equation (Eq. (\ref{Eq:phaseEq})) and the pressure equation. This imposes a limit on the time-step size by introducing a capillary wave time scale described by the Brackbill conditions \cite{1992-Brackbill}. 

In \textit{interReactiveTransferFoam}, the concentration equation (Eq. (\ref{Eq:CST})) is solved sequentially after the PISO loop. The interfacial mass transfer (Eq. (\ref{Eq:mx})) is then computed and re-injected in the continuity (Eq. (\ref{Eq:cont})) and phase equations (Eq. (\ref{Eq:phaseEq})). The space discretization of the convection terms is then performed using the second-order \textit{vanLeer} scheme \cite{1974-vanLeer}. For the compression terms, the interpolation of $\alpha_d\alpha_c$ is carried out using the \textit{interfaceCompression} scheme \cite{2016-OpenFOAM}. The diffusion term $\nabla.\left({D}_j\nabla c_j\right)$ is discretized
using the Gauss linear limited corrected scheme, which is second order and conservative. The discretization of the CST flux is performed using the phase upwinding scheme \cite{2020-Patsoukis}. Finally, the chemical reaction step is solved using Phreeqc \cite{2013-Pakhurst}.

\section{Verification}
The multiphase transport solver has previously been validated by comparison with analytical and semi-analytical solution for a range of 1D, 2D and 3D problems \cite{2018-Maes,2018b-Maes,2020-Maes}. In particular the calculation of the local volume change induced by interface transfer for a soluble phase has been validated by comparison with the analytical solution for dissolution of a gas phase in water in a 1D domain. In this study, we present the validation of the coupling between the multiphase transport and  chemical reactions. 

\label{sec:2}
\subsection{Multiphase reactive transport in 1D at equilibrium}

\begin{figure}[!b]
\begin{center}
\includegraphics[width=0.99\textwidth]{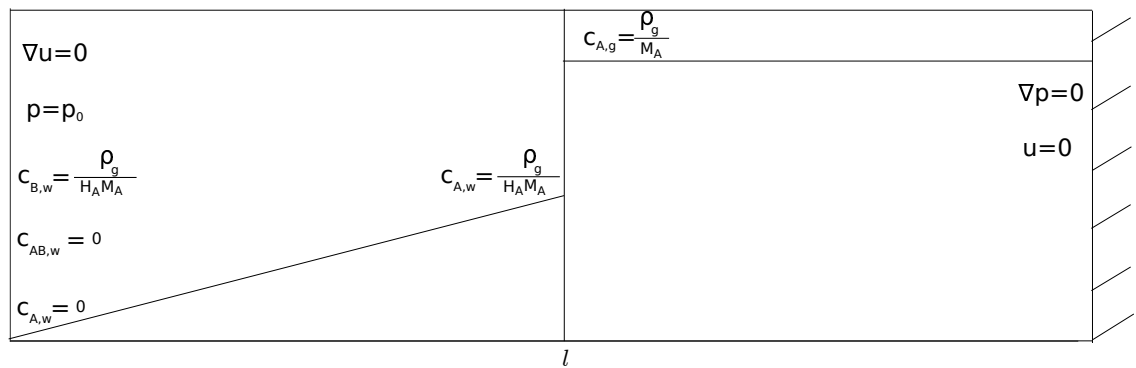}
\caption{Set-up for multiphase reactive transport in 1D at equilibrium  \label{fig:1Dgeometry}}
\end{center} 
\end{figure}

The objective of this test case is to validate the coupling between multiphase transport and chemical reactions by comparison with a system where an analytical solution exists. For this, we consider a system with 3 components (A, B, and AB) and two phases (water and gas). The three component species are diluted in the water phase ($\rho_w=\unit[1000]{kg/m^3}$) with diffusion coefficient all equal to $D=10^{-9}$~m$^2$/s. The gas phase ($\rho_g=\unit[1]{kg/m^3}$) is a pure mixture made of component A ($H_A=10$, $M_{A}=1$ kg/kmol) while B and AB do not cross the interface ($H_B=0$, $M_{B}=1$ kg/kmol and $H_{AB}$, $M_{AB}=1$ kg/kmol). The components in the water phase react following the bimolecular reaction
\begin{equation}\label{Eq:reacAB}
    AB \rightarrow A + B.
\end{equation}
We assume that for this case that all activity coefficient $\gamma_k=1.0$. Therefore, the law of mass action can be written as
\begin{equation}
    c_{AB,w}=\frac{c_{A,w}c_{B,w}}{Kc_0},
\end{equation}
where $K=10.0$ is the equilibrium constant of the reaction (Eq. (\ref{Eq:reacAB})).

The domain is a 1D tube of 1mm length (Fig. \ref{fig:1Dgeometry}).
The gas/liquid interface is initially positioned at a distance
$l_0=0.5$ mm from the left boundary. The left boundary has a constant pressure $p=p_0$, with constant concentration $c_{A,w}=0$, $c_{B,w}=\frac{\rho_g}{H_AM_{A}}$ and $c_{AB,w}=0$, while the right boundary has a no-flow condition.

Since the right boundary has a no-flow condition, and because the fluids are assumed incompressible, the velocity in the gas phase is equal to 0. Hence, the total mass conservation at the interface (Eq. (\ref{interfaceBC3})) can be written as
\begin{equation}
    \rho_w\left(u_w-w\right)=-\rho_g,
\end{equation}
which leads to $u_w\approx w$. Assuming that advective transport is negligible by comparison to diffusive transport, i.e.
\begin{equation}\label{Eq:Peclet}
    Pe=\frac{wl_0}{D}<<1,
\end{equation}
the transport equation (Eq. (\ref{Eq:transport}) can be considered to be at equilibrium at the time-scale of interface displacement. Therefore
\begin{linenomath}
\begin{align}
    D\nabla^2 c_{A,w}+D\nabla^2 c_{AB,w} = 0, \\
    D\nabla^2 c_{B,w}+D\nabla^2 c_{AB,w} = 0.
\end{align}
\end{linenomath}
Since $K>>1$, $c_{AB,w}<<c_{A,w}$ and $c_{AB,w}<<c_{B,w}$, an approximated analytical solution for the concentration in the water phase is
\begin{linenomath}
\begin{align}
    &c_{A,w}=\frac{\rho_g}{M_{A}H_A}\frac{x}{l}, \\
    &c_{B,w}=\frac{\rho_g}{M_{A}H_A}\left(1-\frac{\rho_g}{Kc_0M_{A}H_A}\frac{x}{l}\right),\\
    &c_{AB,w}=\frac{\rho_g^2}{Kc_0M_A^2H_A^2}\frac{x}{l}\left(1-\frac{\rho_g}{Kc_0M_AH_A}\frac{x}{l}\right),
\end{align}
\end{linenomath}
As only the component $A$ crosses the interface,
\begin{equation}\label{Eq:solution1}
    w=M_A\frac{D\nabla c_{A,w}(x=l)}{\rho_g}=\frac{D}{H_Al},
\end{equation}
which shows that Eq. (\ref{Eq:Peclet}) is valid for $H_A>>1$. Finally, integrating Eq. (\ref{Eq:solution1}) gives
\begin{equation}
    l(t)=l_0\sqrt{1+\frac{2Dt}{H_Al_0^2}}.
\end{equation}

\begin{figure}[!t]
\begin{center}
\includegraphics[width=0.99\textwidth]{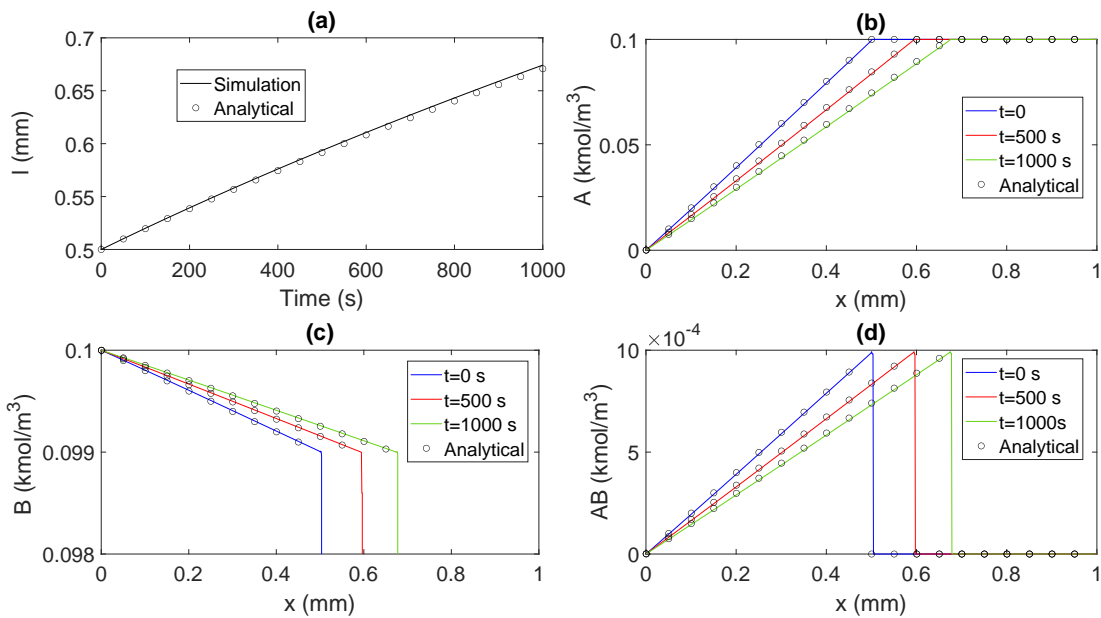}
\caption{Comparison between simulated and analytical results during multiphase reactive transfer in a 1D geometry (see fig. \ref{fig:1Dgeometry}). (a) Evolution of the interface position; (b) Concentration profile of A at different times; (d) Concentration profile of B at different times; (d) Concentration profile of AB at different time. \label{fig:1DReactiveTransferGasWater}}
\end{center} 
\end{figure}

The test case is simulated on a regular grid with 1000 grid blocks, and with a constant time-step t=0.01 s. In order to compare with the analytical solution, the local volume change is initially turned off and the concentration of A in the gas phase is kept equal to 1 kmol/m$^3$ until the concentrations in the water phase reaches an equilibrium. Local volume change is then turned on and the simulation is run until t=1000 s.   

Figure \ref{fig:1DReactiveTransferGasWater} show a comparison between simulated and analytical results. We obtain a very good agreement between the model and the analytical solution, and have thus validated the coupling between multiphase flow with interface transfer and chemical reactions in our model. 

\subsection{Injection of a CaCl solution in an oil-filled tube in 2D}
 \label{sec:3-2}
The objective of this test case is to show that, unlike the CST method, the simplified model (Eq. (\ref{Eq:simplified})) generates artificial mass transfer that damages the numerical solution. First, we consider a 2D straight microchannel of size 300 $\mu$m$\times$ 100 $\mu$m. The fluid properties are summarized in Table \ref{Table:CaCl}. The channel is initially filled with oil. At t=0, we start injecting an aqueous solution of 1000 mg/L of CaCl from the left boundary at velocity $U=3$ mm/s, which corresponds to $Re=0.3$ and $Ca=10^{-4}$. The solid boundaries are assumed to be oil-wet, with a contact angle of 45$^o$. In addition, surface complexation occurs at the surface of the solid following the Na-montmorillonite SCM proposed by Bradbury and Bayens \cite{1997-Bradbury}, which is summarised in Table \ref{chemmodel2}. The surface density $\Gamma$ of adsorption sites $>$S is equal to 2.4 $\mu$mol/m$^2$.

\begin{table}[!ht]
\centering
\begin{tabular}{cccccc}
 & Density (kg/m$^3$)  & Dynamic viscosity (mPa.s) & Interfacial tension (mN/m) \\
\hline
Oil & 864 & 14.3 &  &  \\[0.1cm]
Aqueous solution & 1000& 1 & 30 \\[0.1cm]
\hline
\end{tabular}
\caption{Fluid properties for oil and CaCl solution system.\label{Table:CaCl}}
\end{table}

\begin{table}[!ht]
\centering
\begin{tabular}{ccc}		
No & Surface reactions & $K^i$ \\[0.1cm]
\hline
1 & $>$SOH$^0$ + H$^+$ $\Leftrightarrow$ $>$SOH$_2^+$ & $10^{4.5}$ \\[0.1cm]
2 & $>$SOH$^0$ $\Leftrightarrow$ $>$SO$^-$ + H$^+$ & $10^{-7.9}$ \\[0.1cm]
3 & $>$SOH$^0$+Ca$^{2+}$ $\Leftrightarrow$ $>$SOCa$^+$ + H$^+$ & $10^{-5.9}$ \\[0.1cm]
\hline
\end{tabular}
\caption{Surface-complexation reactions and their intrinsic stability constant on a clay surface \cite{1997-Bradbury}. \label{chemmodel2}}
\end{table} 

The aqueous solution includes 4 dilute species (Ca$^{+2}$, Cl$^{-}$, H$^+$ and OH$^-$). Each of these species only exists in the water phase, so that $H_k=0$ and $D_{k,2}=0$. The diffusion coefficient of species in the water phase are obtained from Li and Gregory \cite{1973-Li} and are summarised in Table \ref{TableDiffu}. The transport of these species in the domain is strongly advection-dominated, with P\'eclet numbers varying from 10.2 to 127.  

We assume that the surface of the solid has been previously equilibrated with the same solution of 1000 mg/L of CaCl. Therefore, the chemical equilibrium should be unchanged and the concentration in the water phase constant.

\begin{figure}[!b]
\begin{center}
\includegraphics[width=0.99\textwidth]{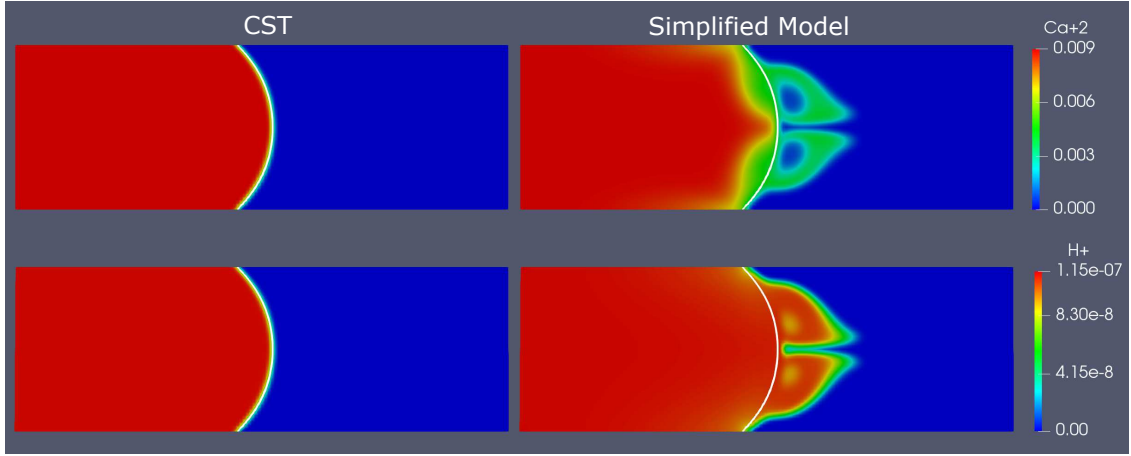}
\caption{Concentration maps for Ca$^{+2}$ and H$^+$ obtained with the CST method and with the simplified method at t=0.15 s. \label{fig:CSTvsSM}}
\end{center} 
\end{figure}

The simulations are performed on a 150$\times$50 cartesian grid with a constant time-step $\Delta t=~0.5$~ms. Figure \ref{fig:CSTvsSM} show the concentration maps for Ca$^{+2}$ and H$^+$ obtained with each method at t=0.15 s. We see that the CST method leads to a sharp interface between species concentration and oil phase fraction, with constant concentration in the aqueous phase. No artificial mass transfer occur and the system remains at chemical equilibrium. However, the simplified method leads to a large amount of artificial mass transfer. The species concentrations in the aqueous phase appear diffused and we obtain significant concentration in the oil phase that is purely induced by numerical errors. Note that the simplified model only considers the concentration in the aqueous phase, so the error of concentration in the oil phase can be ignored. However, as a result of the concentration diffusion in the water phase, the chemical equilibrium is disturbed and the concentration of surface species on the solid boundary changes.

\begin{figure}[!t]
\begin{center}
\includegraphics[width=0.9\textwidth]{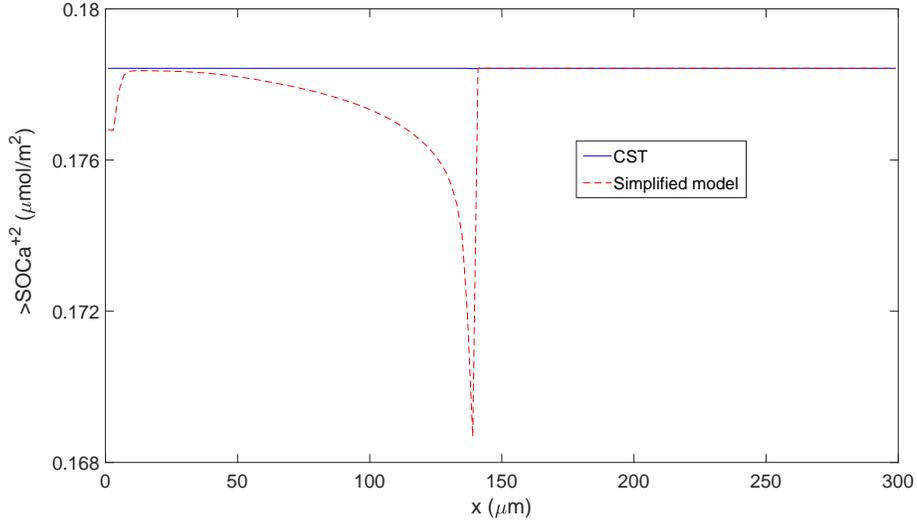}
\caption{Concentration of $>$Ca$^+$ along the x-axis obtained with the CST method and with the simplified method at t=0.15 s. \label{fig:SurfCa}}
\end{center} 
\end{figure}

Figure \ref{fig:SurfCa} shows the concentration of $>$SOCa$^+$ along the x-axis at t=0.15 s. We observe that the CST method leads to a constant concentration with no change of concentration by chemical reaction, while the simplified model has a decrease of 5$\%$ of $>$SOCa$^+$  across the interface, indicating that changes of concentration by chemical reaction have occurred. 

This example demonstrates that the CST method rather than the simplified model should be employed to simulate multicomponent reactive transport in pore-scale images. Additionally, the CST method only requires the computational of two additional fluxes (species compression and CST fluxes), so the increase in CPU time is very limited. For the case presented here, the simplified model ran for 6067 s with 2 processors on an intel Xeon core, while the CST method ran for 6127~s, representing an increase in computational expense of 1$\%$. 

\section{Applications}
In this section we show how GeoChemFoam can be used to simulate and upscale various reactive processes in pore-scale geometries.

\subsection{Test Case 1: CO$_2$ gas dissolution in a 3D pore cavity}
\label{secCO2}
\begin{figure}[!b]
\begin{center}
\includegraphics[width=0.7\textwidth]{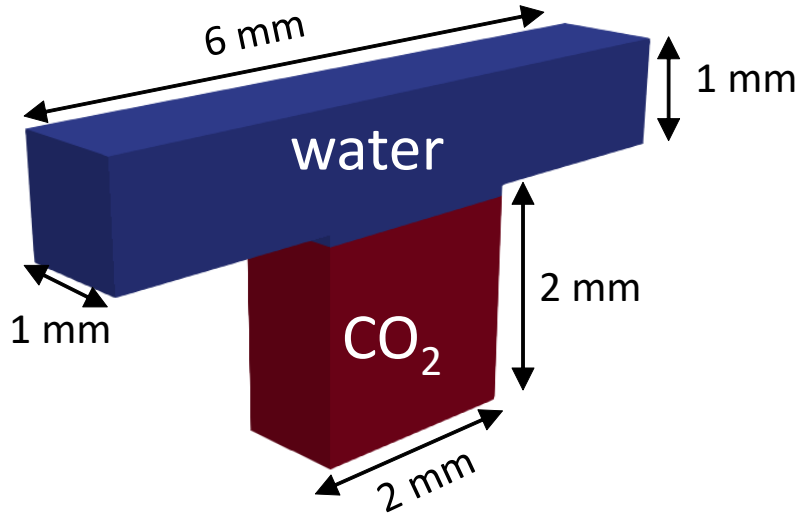}
\caption{Schematic diagram of the cavity geometry and initial conditions (from \cite{2020-Patsoukis}\label{fig:cavityGeo}}
\end{center} 
\end{figure}

In this example, we investigate interface transfer and chemical reactions during dissolution of a CO$_2$ gas bubble in a pore cavity. The model domain is the same as presented in \cite{2020-Patsoukis}. The geometry is a 6mm$\times$1mm$\times$1mm channel, with a 2mm$\times$2mm$\times$1mm cavity inserted in the middle (Fig. \ref{fig:cavityGeo}). The domain is meshed using a uniform grid with resolution 50 microns. Initially,
CO$_2$ gas is trapped in the cavity and the rest is filled with water. The fluid properties are summarized in Table \ref{Table:cavity}.

\begin{table*}[!ht]
\centering
\begin{tabular}{cccccc}
 & Density (kg/m$^3$)  & Dynamic viscosity (mPa.s) & Interfacial tension (mN/m) \\
\hline
Gas & 1.87 & 1.496$\times 10^{-2}$ &  &  \\[0.1cm]
Water & 1000& 1 & 50 \\[0.1cm]
\hline
\end{tabular}
\caption{Fluid properties for CO$_2$ dissolution in a cavity.\label{Table:cavity}}
\end{table*}

The system contains 6 species (H$_2$O, H$^+$, OH$^-$, CO$_2$, CO$_3^{2-}$ and HCO$_3^-$). Each species with the exception of H$_2$O is dilute in the aqueous phase. The gas phase is pure CO$_2$. The diffusion coefficient and Henry's constant are summarized in Table \ref{TableDiffu}.

The system includes three chemical reactions that are summarized in Table \ref{Table:reactionCO2}. As CO$_2$ dissolves in the water phase, H$^+$ and HCO$_3^-$ are created and the chemical equilibrium is modified, leading to a decrease in pH.

\begin{table*}[!ht]
\centering
\begin{tabular}{cc}
 Reaction & K \\[0.1cm]
\hline
H$_2$O $\rightleftharpoons$ H$^+$ + OH$^{-}$  &   $K_1=1.01\times10^{-14}$ \\[0.1cm]
HCO$_3^-$ $\rightleftharpoons$ H$^{+}$ + CO$_3^{2+}$ & $K_2=4.9\times10^{-11}$\\[0.1cm]
CO$_2$+H$_2$O $\rightleftharpoons$ H$^+$ + HCO$_3^{-}$  & $K_3=4.5\times10^{-7}$   \\[0.1cm]
\hline
\end{tabular}
\caption{CO$_2$-water reactions.\label{Table:reactionCO2}}
\end{table*}

At t=0, we inject pure water at pH=7 from the left boundary at a flow rate of 0.1 mL/min which corresponds to a capillary number of 3.3$\times10^{-6}$. The simulation is run until t=3 min with a constant time-step $\Delta t=20$ $\mu$s.

\begin{figure}[!b]
\begin{center}
\includegraphics[width=0.99\textwidth]{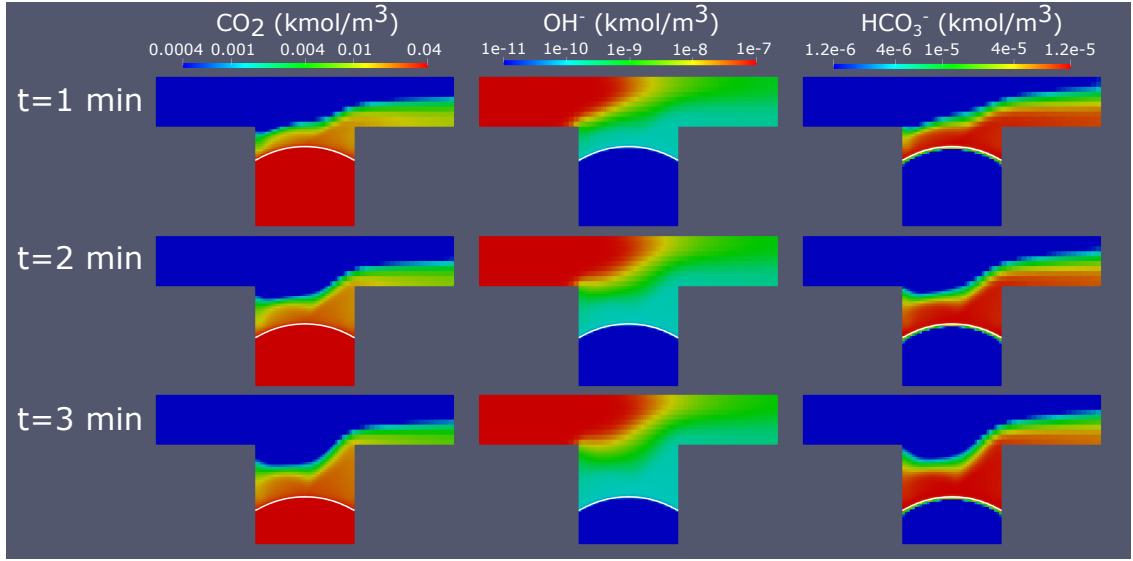}
\caption{Concentration map of CO$_2$, OH$^-$ and HCO$_3^-$ at the mid-plan during dissolution of a CO$_2$ bubble in a 3D pore cavity at t= 1 min, 2 min and 3 min. The gas/water interface is shown in white and the concentration are shown with a color map on a log scale to enhance the contrasts\label{fig:cavityConcentration}}
\end{center} 
\end{figure}

Fig. \ref{fig:cavityConcentration} show the concentration map of CO$_2$, OH$^-$ and HCO$_3^-$ at the mid-plane at t= 1 min, 2 min and 3 min. The gas/water interface is shown in white. The concentrations are shown with a color map on a log scale to enhance the contrast. We observe that the mixing of species in the water phase is poor. This is because, even though the flow rate is low with $Re$ and $Ca$ well into the creeping and capillary dominated regime, the transport of species is still advection-dominated. For example, the P\'eclet number for the CO$_2$ species is equal to 104. Therefore, there is a strong difference between the concentrations upstream and downstream of the cavity. From the inlet and up to the cavity, pH is close to 7 with no CO$_2$ present. Within the cavity, the water on top of the gas bubble has a pH close to 4 and a CO$_2$ concentration close to 0.03 kmol/m$^3$. From the end of the cavity to the outlet, the pH is close to 5 and CO$_2$ is present at the bottom of the channel with a concentration close to 0.004 kmol/m$^3$, but no CO$_2$ is present in the top part of the channel.

\begin{figure}[!t]
\begin{center}
\includegraphics[width=0.99\textwidth]{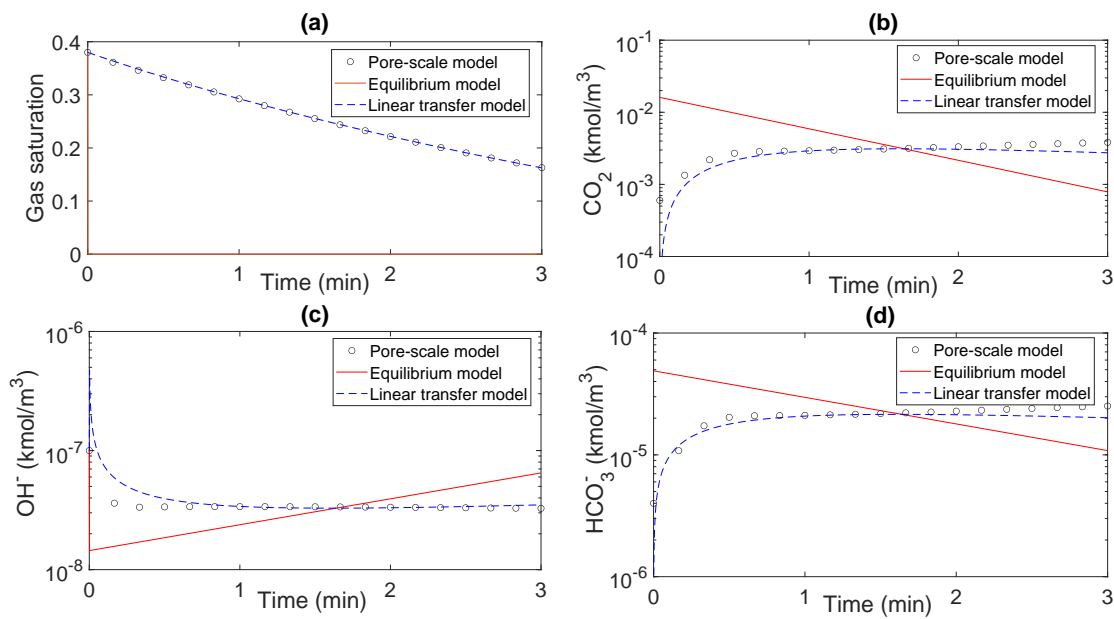}
\caption{Evolution of (a) gas saturation, and of concentration of (b) CO$_2$, (c) OH$^-$ and (d) HCO$_3^-$ in the water phase obtained with pore-scale, equilibrium and linear transfer with mixing-induced reaction rates models during dissolution of a gas bubble in a 3D pore cavity. \label{fig:cavityUpscaled}}
\end{center} 
\end{figure}

Poor mixing has a strong impact when upscaling the chemical reactions to the larger scales. In Fig. \ref{fig:cavityUpscaled}, the evolution of gas saturation as well as the concentrations of CO$_2$, OH$^-$ and HCO$_3^-$ obtained in the pore-scale simulation are compared with the results obtained when using a fully-mixed equilibrium model. The results diverge significantly as the concentrations in the equilibrium model trend in the opposite direction to those of the pore-scale simulation. This divergence occurs with the equilibrium model because the CO$_2$ dissolves instantaneously in the water phase, forming a carbonic acid that significantly reduces the pH of the water, and then the acid is slowly flushed out of the domain and the water becomes increasingly neutral. 

\begin{figure}[!b]
\begin{center}
\includegraphics[width=0.9\textwidth]{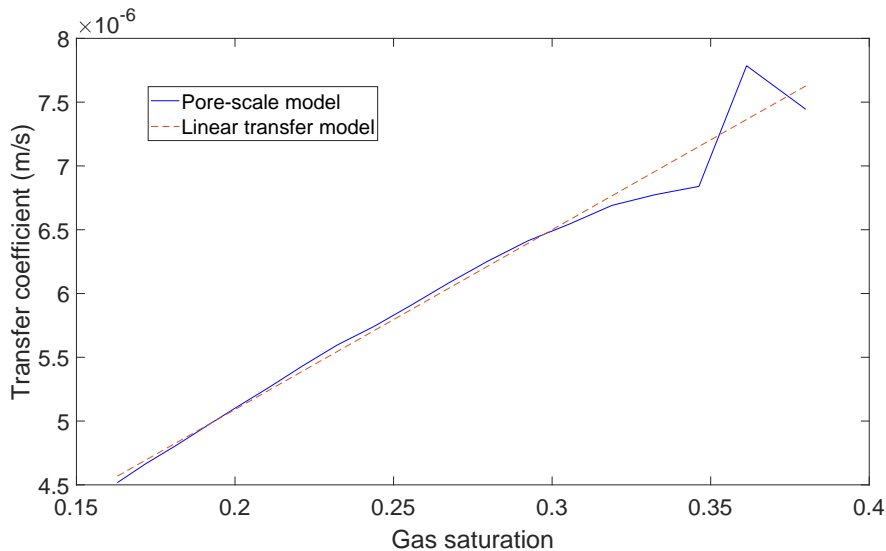}
\caption{Mass exchange coefficient calculated from the pore-scale simulation during dissolution of a CO$_2$ gas bubble in a 3D pore cavity and linear approximation  $\lambda_{\text{CO}_2}\approx\lambda_{\text{CO}_2}^o+\lambda_{\text{CO}_2}^1S_2$ use in the linear transfer model, with $\lambda_{\text{CO}_2}^o=10^{-4}$ m/s and $\lambda_{\text{CO}_2}^1=6.2\times10^{-4}$ m/s\label{fig:cavityTransferCoeff}}
\end{center} 
\end{figure}
However, in reality, the phase transfer occurs on a much larger time-scale (Fig. \ref{fig:cavityUpscaled}a) and thus a linear transfer model would be more appropriate to simulate this at the larger scale. Using Equ. (\ref{Eq:k}), the mass exchange coefficient for CO$_2$ can be calculated from the results of the pore-scale simulation as
\begin{equation}
    \lambda_{\text{CO}_2}= \frac{M_{\text{CO}_2}}{A_{12}\left(H_{\text{CO}_2}C_{\text{CO}_2,1}-C_{\text{CO}_2,2}\right)}.
\end{equation}
The mass exchange coefficient is plotted as a function of the gas saturation $S_2$ in Fig. \ref{fig:cavityTransferCoeff} and we observe that it can be approximated as a linear function of $S_2$
\begin{equation}
    \lambda_{\text{CO}_2}\approx\lambda_{\text{CO}_2}^o+\lambda_{\text{CO}_2}^1S_2
\end{equation}
where $\lambda_{\text{CO}_2}^o=10^{-4}$ m/s and $\lambda_{\text{CO}_2}^1=6.2\times10^{-4}$ m/s. The evolution of the gas saturation in the domain can then be estimated using this linear transfer model, and the results are plotted on Fig. \ref{fig:cavityUpscaled} and compared to the pore-scale and equilibrium models. Contrary to the equilibrium model, the evolution of saturation obtained using the linear transfer model are well-fitted to the results of the pore-scale simulations.  

\begin{figure}[!b]
\begin{center}
\includegraphics[width=0.99\textwidth]{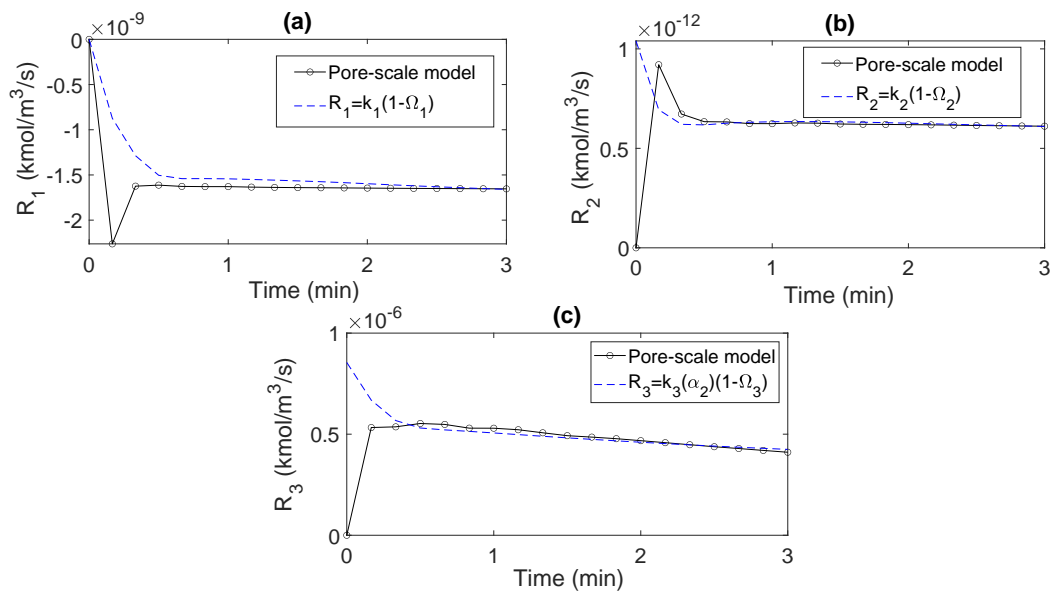}
\caption{Evolution of the reaction rates of the three reactions present in a CO$_2$ water system (Table \ref{Table:reactionCO2}) during dissolution of a CO$_2$ gas bubble in a 3D pore cavity, and comparison with mixing-induced reaction models with $k_1=1.80\times10^{-11}$, $k_2=1.04\times10^{-12}$ and $k_3=\frac{5.30\times10^{-7}}{1-\alpha_2}$ kmol/m$^3$/s.\label{fig:cavityMixing}}
\end{center} 
\end{figure}

In addition, the incomplete mixing in the water phase induces a delay in the chemical reactions and the phase average concentration of species in the domain are not at chemical equilibrium. Mixing-induced reaction rates can be calculated during the pore-scale simulation by integrating the changes of concentrations obtained by chemical reaction (calculated by Phreeqc) over the full simulation domain. Fig. \ref{fig:cavityMixing} shows the evolution of the reaction rates of the three reactions present in the system (Table \ref{Table:reactionCO2}). We observe that the rates of reactions 1 and 2 converge toward a plateau, which is typical of a mixing-induced reaction constant that does not depend on saturation. However, the rate of reaction 3 consistently decreases from t=0.5 min, which suggests that the mixing-induced reaction constant $k_3$ decreases as the gas saturation increases.  The saturation indexes $\Omega_1$, $\Omega_2$ and $\Omega_3$ are calculated based on the averaged concentrations in the water obtained from the pore-scale simulation, and the mixing-induced reaction rates are calculated with constant $k_1=1.80\times10^{-11}$ and $k_2=1.04\times10^{-12}$ kmol/m$^3$/s. These along with $k_3=\frac{5.30\times10^{-7}}{1-\alpha_2}$ kmol/m$^3$/s are plotted in Fig. \ref{fig:cavityMixing} and compared with the rates obtained from the pore-scale simulation results. We observe that the mixing-induced rates are well fitted to the pore-scale simulation results after an initialisation time of about 0.5 min. These mixing-induced rates are included in the linear transfer model, and the concentration of CO$_2$, OH$^-$ and HCO$_3$ obtained are plotted in Fig. \ref{fig:cavityUpscaled} and compared to the results of the pore-scale and equilibrium models. Contrary to the equilibrium model, the evolution of the average concentrations in the water phase obtained using the linear transfer model are well-fitted to the results of the pore-scale simulations. We can thus analyse the results of the pore-scale simulation to develop an accurate upscaled model based on linear transfer and mixing-induced reaction rates.   

\subsection{Test Case 2: Injection of a CaCl solution in a micro-CT image}

\begin{figure}[!t]
\begin{center}
\includegraphics[width=0.99\textwidth]{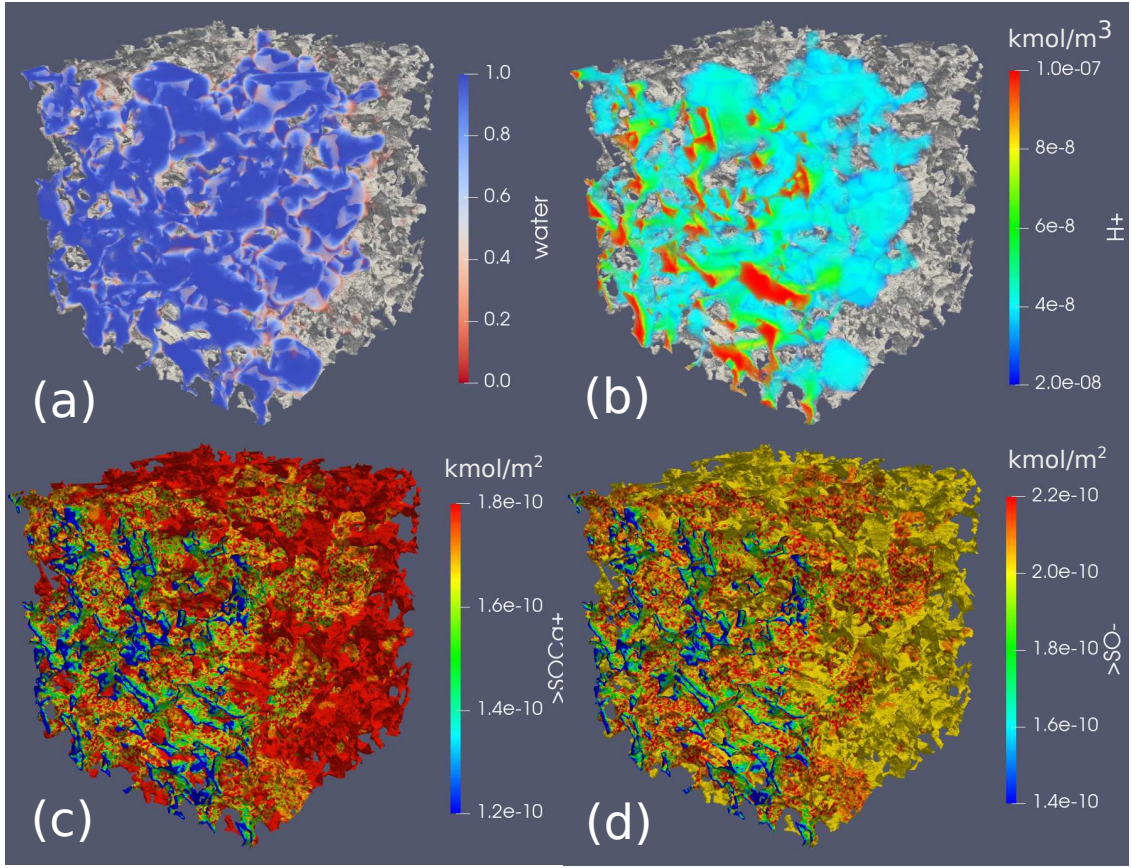}
\caption{(a) Water phase fraction, (b) concentration of H$^+$ in the water in the bulk, and (c) concentration of $>$SOCa$^+$ and (d) $>$SO$^-$ on the solid surfaces at t=0.5 during injection of a CaCl solution in a micro-CT image of Bentheimer sandstone.\label{fig:Bentheimer}}
\end{center} 
\end{figure}

We now investigate multiphase multicomponent reactive transport in a micro-CT image. First we simulate aqueous CaCl injection into an oil saturated pore space with surface complexation. Then from the pore scale result we calculate volume averaged saturation and concentration and compared it to the result an upscaled equilibrium model. Then we propose a correction to the upscaled equilibrium model based on a reduced surface charge. 

The image is a 1000$^3$ voxel micro-CT image of Bentheimer sandstone with a resolution of 2.5 microns, which can be downloaded from the Digital Rock Portal \href{https://dx.doi.org/10.17612/f4h1-w124}{https://dx.doi.org/10.17612/f4h1-w124}. A 512$^3$ voxel image is extracted from the center of the image for this example. 

The domain is meshed using the OpenFoam\textsuperscript{\textregistered} \textit{snappyHexMesh} utility \cite{2016-OpenFOAM}. First a 128$^3$ cartesian grid is generated. Next, each grid block that is crossed by the solid surface is refined once in each direction, leading to resolution of 5 microns. The cells in the solid phase are then removed, while the cells that intersect the rock/pore interface are replaced by hexahedral or tetrahedral cells that match the solid boundaries. The final mesh contain 2,315,379 cells. The porosity $\phi$ can then be calculated from the mesh and the absolute permeability $K_a$ can be estimated by solving the Stokes equation \cite{2012-Talon}. Our image has a porosity of 0.22 and a permeability of  $2.9\times10^{-12}$ m$^2$ .

The fluid properties (Table \ref{Table:CaCl}) and  chemical system (Table \ref{chemmodel2}) are the same as the ones used in Section \ref{sec:3-2}. The pore space is initially filled with oil and the surface of the solid has been previously equilibrated with a solution of 1000 mg/L of CaCl. At t=0, we inject from the left boundary with a solution of 100 mg/L of CaCl at constant velocity $U$=3mm/s, corresponding to a capillary number $Ca=10^{-4}$. A constant pressure is set at the right boundary, while the top, bottom, front and back boundaries have a no-flow condition. The solid boundaries are assumed to be oil-wet, with a contact angle of 45$^o$. The simulation is run until t=0.5 s with a constant time-step $\Delta t=1$ $\mu$s with 24 processors on an intel Xeon core. The total CPU time of this simulation was 31 days.

Fig. \ref{fig:Bentheimer} shows the water phase fraction, the concentration of H$^+$ in the water in the bulk phase, and the concentration of $>$SOCa$^+$ and $>$SO$^-$ on the solid surfaces at t=0.5 s. Although the mixing of H$^+$ is not complete, it is better than the mixing in the previous test case, with most values of H$^+$ concentration close to 4$\times10^{-8}$ kmol/m$^3$. However, the mixing on the solid surface is very poor.

\begin{figure}[!t]
\begin{center}
\includegraphics[width=0.8\textwidth]{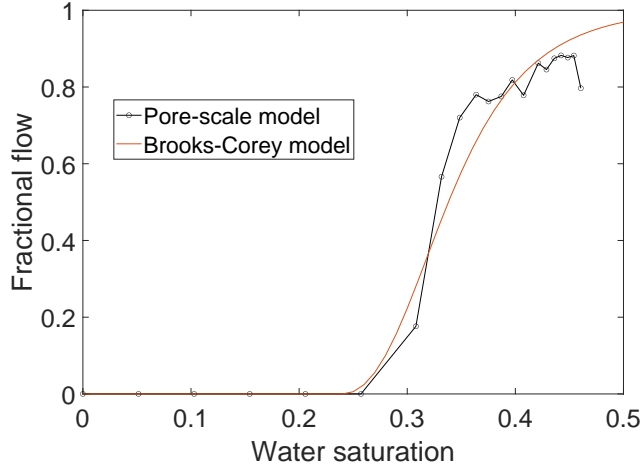}
\caption{Fractional water flow as a function of water saturation during injection of a CaCl solution in a micro-CT image of Bentheimer sandstone, calculated from the pore-scale simulation results and using the Brook-Corey model, with $k_{r1,max}=k_{r2,max}=1.0$, $S_{wc}=0.24$, $S_{nar}=0.25$, $n_1=2$ and $n_2=3$. \label{fig:FracFlow}}
\end{center} 
\end{figure}

\begin{figure}[!t]
\begin{center}
\includegraphics[width=0.99\textwidth]{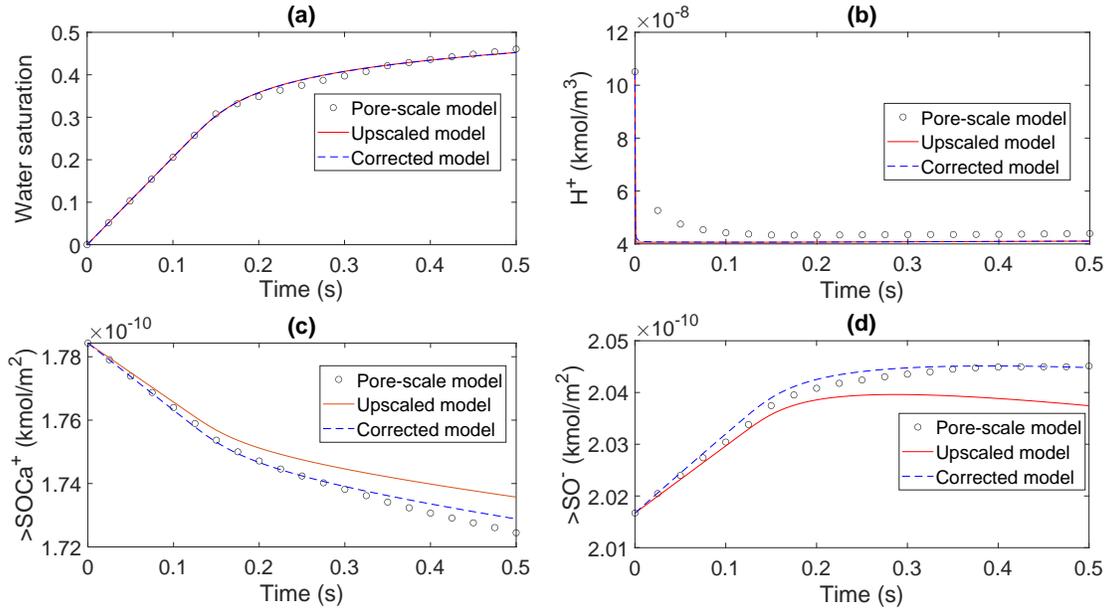}
\caption{Evolution of (a) water saturation, (b) phase averaged concentration of H$^+$ in the water, and (c) average concentration of $>$SOCa$^+$ and (d) $>$SO$^-$ on the solid surfaces obtained using a pore-scale model, an upscaled model and a corrected upscaled model during injection of a CaCl solution in a micro-CT image of Bentheimer sandstone. \label{fig:BentheimerUpscaled}}
\end{center} 
\end{figure}

The fractional flow of water at the outlet is calculated from the pore-scale results and plotted in 
Fig. \ref{fig:FracFlow}. The curve is fitted to a Brooks-Corey model where $k_{r1,max}=k_{r2,max}=1.0$, $S_{wc}=0.24$, $S_{nar}=0.25$, $n_1=2$, and $n_2=3$, which is also plotted on Fig. \ref{fig:FracFlow}. Fractional flow is used in an upscaled model to calculate the evolution of the total water saturation in the domain. The results are plotted in Fig. \ref{fig:BentheimerUpscaled}a along with the water saturation obtained with the pore-scale simulation. The upscaled model fits the pore-scale simulation with a high degree of accuracy.

We then run a reactive transport model using an upscaled equilibrium model, where the mass action laws (Eq. (\ref{Eq:massaction}) are calculated using the average concentration of solution species in the water phase and the average concentration of surface species on the solid surface. The results are plotted in Fig. \ref{fig:BentheimerUpscaled}b, c and d, along with the concentrations obtained
in the pore-scale simulation. We observe that the equilibrium model predicts a higher concentration of $>$SOCa$^+$ and a lower concentration of $>$SO$^-$. This suggests that the equilibrium model does not overpredict the reaction rate, like in the previous case, but underpredicts it. Therefore, the model cannot be improved by defining mixing-induced reaction rates. Instead, the chemical equilibrium itself should be modified. Since the model predicts a higher concentration of $>$SOCa$^+$ and a lower concentration of $>$SO$^-$, the surface charge in the equilibrium model is lower in absolute value than the one obtained in the pore-scale model. This discrepancy will have a large impact on the chemical equilibrium as the equilibrium constant depends strongly on the surface charge through the surface potential (Eq. (\ref{Eq:Ki})). 

In order to obtain a more accurate prediction, the model is corrected by multiplying the surface charge $q$ (Eq. (\ref{sigma})) by 0.95 before calculating the surface potential (Eq. (\ref{Grahame})). The results of the corrected model are plotted in Fig. \ref{fig:BentheimerUpscaled}. The corrected model gives significantly more accurate results than the initial upscaled model. However, the errors in the surface concentrations are increasing and the concentration of H$^+$ in the bulk remains significantly lower than the one obtained in the pore-scale simulation. This suggests that the model could be further improved by defining mixing-induced reaction rates with the corrected equilibrium constant.

\section{Conclusion}
In this study, we presented a novel multiphase reactive transport model to perform direct numerical
simulation of multiphase flow, multicomponent transport and geochemical reactions on pore space images. We built  a model GeoChemFoam, which is based on OpenFOAM\textsuperscript{\textregistered} \cite{2016-OpenFOAM}, an established library to solve partial differential equations, and Phreeqc \cite{2013-Pakhurst}, the most prevalent geochemical solver. The multiphase flow was solved using the VOF method \cite{1981-Hirt}, and the transport of species using the CST method \cite{2018-Maes}. The reactive transport solver was based on a sequential non-iterative operator splitting approach \cite{2004-Carrayrou} and the chemical equilibrium was solved with Phreeqc.

The method was validated successfully for simple configurations where analytical
solutions exist. In particular, we showed that the CST method provides an accurate representation of interface boundary conditions free of artificial mass transfer, and it can therefore be applied to model reactive transport in multiphase systems.

We then applied the model to two test cases. In test case 1, we simulated reactive transport during dissolution of a CO$_2$ gas bubble in a 3D pore cavity. The liquid/gas interface was tracked as well as the concentration of each reactive species in the domain and incomplete mixing was observed. We showed that an upscaled model based on phase and chemical equilibrium could not predict accurately the evolution of average phase saturations and species concentrations in the domain. Instead, the total flux of interface transfer and the average reaction rates in the domain were calculated and we showed that an upscaled model based on linear transfer and mixing-induced reaction rates could accurately predict the evolution of average phase saturations and species concentrations in the domain.

Finally, in test case 2 the model was applied to simulate multiphase reactive transport in a micro-CT image of Bentheimer sandstone where a solution of CaCl was injected into an oil saturated domain with surface complexation at the solid surface. The concentration map of each species on the solid surface was calculated and we observed a poor mixing of charge on the surface. We then ran an upscaled model based on chemical equilibrium and observed that it was overpredicting the change of surface concentration by chemical reactions. Thus we show that surface concentrations cannot be modelled by mixing-induced reaction rates, and the chemical equilibrium need to be modified to take these into account. We then demonstrated that a corrected model that multiply the total surface charge by 0.95 was giving a significantly more accurate result.

The work presented in this paper has wide ranging applications in the oil and gas, carbon capture and storage, contaminant transport, battery, and fuel cell industries. Our simulation framework together with the upscaling methodologies proposed in this
paper are an important step forward in our objective of fully characterizing multiphase reactive transport in porous media. Furthermore, this model enables the use of sensitivity analysis to understand how upscaled properties such as the mass exchange coefficient and mixing-induced reaction rates can change with respect to system properties such as injected flow rate or pore-size distribution. In addition, this numerical model can now be bootstraped to field scale multiphase reactive transport simulators using machine-learning regression models by extending work already done for single-phase flow and transport \cite{2021-Menke} with the ultimate goal of developing upscaling strategies that do not require pore-scale simulations \cite{2007-Lichtner}. 

\section*{Declaration}

\subsection{Funding}

This work was done as part of the UK EPSRC funded project on Direct Numerical Simulation for Additive Manufacturing in Porous Media (grant reference EP/P031307/1).

\subsection{Conflicts of interest/Competing interests}
The authors declare no competing interests

\subsection{Availability of data, code and material}
All data, code and material are available online at www.julienmaes.com/geochemfoam

\bibliographystyle{spphys}       
\bibliography{References}   

%
%

\end{document}